\documentclass[10pt]{iopart}

\usepackage{iopams}
\usepackage{graphicx}
\usepackage{setstack}
\usepackage{dcolumn}
\usepackage{bm}
\usepackage{color}
\usepackage{amsbsy}
\usepackage{subfig}

\newcommand{\vect}[1]{\mathbf{#1}}
      \def\mbf#1{\mathchoice{\hbox{\boldmath $\displaystyle #1$}}
      {\hbox{\boldmath $\textstyle #1$}}
      {\hbox{\boldmath $\scriptstyle #1$}}
      {\hbox{\boldmath $\scriptscriptstyle #1$}}}

\begin{document}

\title[]{Modulation of LISA free-fall orbits due to the Earth-Moon system}

\author{Massimo Cerdonio$^1$, Fabrizio De Marchi$^2$\footnote[7]{Corresponding author: fdemarchi@science.unitn.it}, Roberto De Pietri$^3$, Philippe Jetzer$^4$, Francesco
Marzari$^1$, Giulio Mazzolo$^5$, Antonello Ortolan$^6$ and Mauro Sereno$^{7,8}$}

\address{$^1$ Department of Physics, University of Padova and INFN Padova, via Marzolo 8, I-35131 Padova, Italy}

\address{$^2$ Department of Physics, University of Trento and INFN Trento, I-38100 Povo (Trento), Italy}

\address{$^3$ Department of Physics, University of Parma and INFN Parma I-43100 Parma, Italy}

\address{$^4$ Institute of Theoretical Physics, University of Z\"urich, Winterhurerstrasse 190, 8057 Z\"urich, Switzerland} 

\address{$^5$ Max Planck Institut f\"ur Gravitationsphysik, Callinstrasse 38, 30167 Hannover, Germany}

\address{$^6$ INFN Laboratori Nazionali di Legnaro, Viale dell'Universit\`a  35020 Legnaro (Padova), Italy}

\address{$^7$ Dipartimento di Fisica, Politecnico di Torino, Corso Duca degli Abruzzi 24, 10129 Torino,
Italia} 

\address{$^8$ INFN, Sezione di Torino, Via Pietro Giuria 1, 10125, Torino, Italia} 


\begin{abstract}
We calculate the effect of the Earth-Moon (EM) system on the free-fall motion of LISA test masses. We show that the periodic gravitational pulling of the EM system induces a resonance with fundamental frequency 1 $yr^{-1}$ and a series of periodic perturbations with frequencies equal to integer harmonics of the synodic month ($\simeq 3.92 \times 10^{-7}\ Hz$). We then evaluate the effects of these perturbations (up to the 6$^{th}$ harmonics) on the relative motions between each test masses couple, finding that they range between $3\ mm$ and $10\ pm$ for the 2$^{nd}$ and 6$^{th}$ harmonic, respectively.

\noindent
If we take the LISA sensitivity curve, as extrapolated down to $10^{-6}\ Hz$ in \cite{bender03}, we obtain that a few harmonics of the EM system can be detected in the Doppler data collected by the LISA space mission. This suggests that the EM system gravitational near field could provide an additional crosscheck to the calibration of LISA, as extended to such low frequencies.
\end{abstract}



\maketitle

\section{Introduction}
\label{intro}
LISA (Laser Interferometer Space Antenna) is a ten years long NASA-ESA space mission to detect gravitational waves in the frequency range $10^{-4} - 10^{-1}\ Hz$ \cite{lisa1}. It consists of three spacecrafts whose mutual distances are about~$L=5\times10^6\ km$.
The LISA constellation will orbit around the Sun following the same path of the Earth, $\phi_0= 20^\circ$ behind \cite{lisa2}.

\noindent
The ideal configuration for LISA performances should be a rigid equilater triangle \cite{dhu05}; however, the shape of the LISA constellation is subject to significant variations because of the gravitational interaction due to Sun and planets \cite{dhu08}.

\noindent
Much smaller perturbations can be induced by the presence of interplanetary dust \cite{dust} or dark matter in the solar system \cite{darkmatter}.
\noindent
The perturbations due to each celestial body can be treated, at a first approximation, independently. The gravitational effects are quite different in intensity (orders of magnitude) and/or in behavior (stationary or time-dependent); the frequencies involved are, in general, not commensurable and so resonance effects are not observed.
\noindent
The only significative exception is the Earth perturbation \cite{dhu08}, \cite{Povoleri06} and \cite{dhu09}, which gives a resonance because of the 1:1 commensurability between the Earth and LISA orbits.

\noindent
However, such a resonance will not have enough time to grow during the 10 years of LISA mission.
In this paper we focus on the perturbations induced by the EM system on LISA at the frequency of the synodic month and its harmonics, which are much higher than $1\ yr^{-1}$ and so, in a first order approximation, perturbative effects can be treated independently.
\noindent
We extend the approach of \cite{dhu08} to include the time dependent perturbation of the Moon.
\noindent
The plan of the paper is as follows. In \Sref{dynamics} we shortly describe the perturbative approach used to study the EM effects.
\Sref{ems} is devoted to illustrate the EM system and the approximations we used to
model its gravitational near field. In \Sref{perturb} we calculate the modulations of the distance between two LISA test masses due to EM system.
In \Sref{planets} we estimate the perturbations induced by Venus and Jupiter, compared to those due to the EM system calculated in the previous Sections.
\noindent
In \Sref{conclusion} conclusions are drawn and the future research potential of method for further studies and application to LISA are given.

\section{Perturbative dynamics of the LISA test masses}\label{dynamics}
In order to simplify the notation, we use the Astronomical System of Units, for length (AU), mass (M$_\odot$) and time (days). However, the quantities that affect the relative motion of the LISA spacecrafts will be reconverted in the SI units.

\begin{figure*}[h!]
\centering
\includegraphics[width=.55\columnwidth,height=.48\columnwidth]{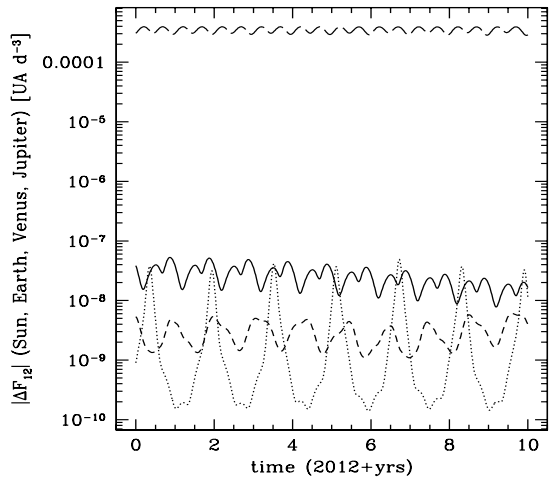}
\caption{
Gravity gradients between two spacecraft due to EM system (solid line), Venus (dotted line), Jupiter (short dashed line) and Sun (long dashed lines).
}
\label{dforce}
\end{figure*}

\noindent
By means of a F77 code based on the inverse 15$^{th}$-order Runge-Kutta method \cite{tj}, we calculate the modulus of the force gradient between 2 LISA spacecrafts due to the main Solar System bodies, i.e. Sun, Venus, Earth (more precisely the EM system) and Jupiter (\Fref{dforce}). The effects are different by orders of magnitude in amplitude and show also different frequencies.

\noindent
In fact, the Earth and Jupiter tidal effects are $10^4$ and $10^5$ times smaller than the Sun contribution, respectively; after 3 years, Sun, Earth and Jupiter cause arms length changes (the so-called ''arm breathing'') of $\simeq 1.2 \times 10^{5}\, km$, 4.8$\times 10^{4}\ km$ and 4$\times 10^{3}\ km$, respectively \cite{dhu05}, \cite{dhu08}.

\noindent
The Venus contribution oscillates in time by two orders of magnitude and it is comparable with Earth effects for a short time interval every 584$\ d$ \cite{dixon}.
However, as we will show in \Sref{planets}, this perturbation is negligible with respect to the Earth one.

\noindent
We therefore conclude that the celestial bodies affect the LISA arms length by at most a few percent. As a consequence, at any time the distance of each satellite from the LISA barycenter is $\rho_0=L/\sqrt{3} \simeq 1.9 \times 10^{-2} \ AU$, and the distance between the Earth and the LISA barycenter is $r_g= 2\, \sin (\phi_0/2) \simeq 0.347 \ AU$, within a relative fluctuation of few percent.

\subsection {Hill-Clohessy-Wiltshire (HCW) reference frame}
\begin{figure*}[h!]
\centering
\includegraphics[width=.65\columnwidth,height=.48\columnwidth]{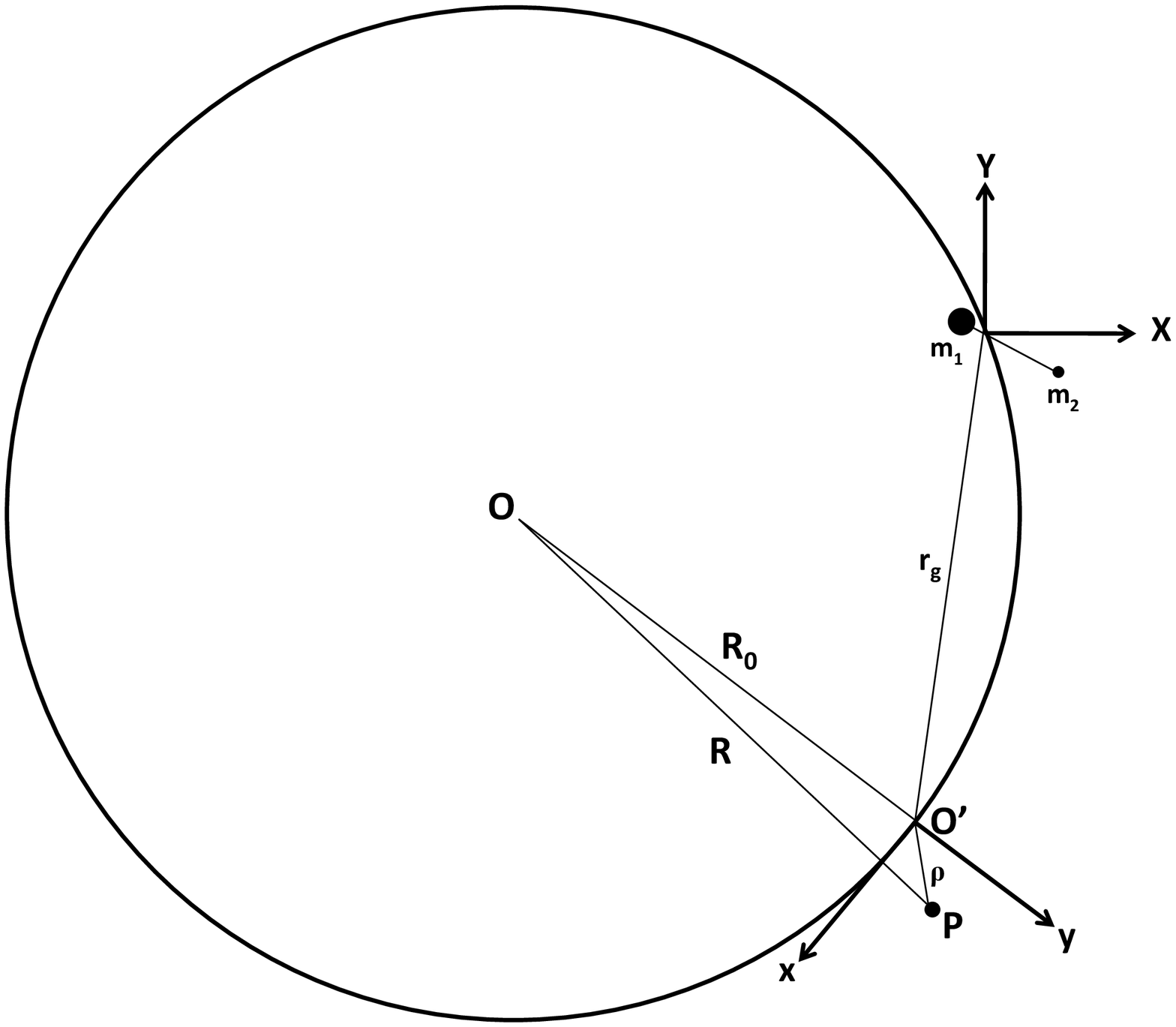}
\caption{ Hill-Clohessy-Wiltshire frame: $R_0$ is the radius of the orbit, $\rho$ and $R$ are the the distances of $P$ from the origin of the rotating and fixed frames, respectively. $\{X,Y,Z\}$ is the associated reference frame (see details in \Sref{ems}). The EM system is also represented. The figure is not to scale.}
\label{cw}
\end{figure*}

To study the EM system effect, we made use of the Hill-Clohessy-Wiltshire (HCW) reference frame {$\{x,y,z\}$} ~\cite{cw60} defined as follows (see \Fref{cw}):

\begin{enumerate}
\item the origin $O'$ describes a circular orbit on the ecliptic plane at $1\ AU$ from the Solar System Barycenter $O$ \cite{ssb};
\item the $xy$ plane coincides with the ecliptic;
\item the $z$-axis is perpendicular to the ecliptic plane and parallel to the Solar System total angular momentum\footnote {The Solar System total angular momentum is perpendicular to a plane slightly inclined with respect to the ecliptic one. For our scopes, such inclination is negligible.};
\item the $x$-axis is tangent to the orbit and is antiparallel to the origin $O'$ velocity vector;
\item the $y$-axis is directed radially outward.
\end{enumerate}
\noindent

\noindent
In this frame the coordinates of the EM system barycenter are $(x_g,y_g,z_g)=R_0(-\sin \phi_0,\cos \phi_0 -1,0)$.
\noindent
Of course, the LISA spacecrafts can be considered as three proof masses.
\noindent
Their equations of motion in the HCW frame read

\begin{equation}
\eqalign{
 \ddot x -2\omega_0 \dot y -\omega_0^2 x =-\frac{\mu}{R^3} x\cr
 \ddot y +2\omega_0 \dot x -\omega_0^2 (y+R_0) =-\frac{\mu}{R^3} (R_0+y)\cr
 \ddot z = -\frac{\mu}{R^3} z\ ,
}\label{eqhcwexact}
\end{equation}

\noindent
where $\mu=G M_\odot =\omega_0^2 R_0^3$, $\omega_0=2 \pi / 365.257\,  d^{-1}$, $R_0=1\,AU$, and $R=\sqrt{x^2+(y+R_0)^2+z^2}$.
\noindent
Being the breathing length $\Delta L \ll L$, each satellite is located at any time at a distance $\rho(t)=\sqrt{x(t)^2+y(t)^2+z(t)^2} \simeq \rho_0$.
\noindent
Since $\rho_0 \ll R_0$, we expand the acceleration due to the Sun in terms of $x/R_0,y/R_0,z/R_0$.
\noindent
Retaining the first term of the series, we have the so-called Hill-Clohessy-Wiltshire equations \cite{cw60} of relative motion

\begin{equation}
\eqalign{
\ddot x -2 \omega_0 \dot y = 0\cr
\ddot y +2 \omega_0 \dot x -3 \omega_0^2 y = 0\cr
\ddot z + \omega_0^2 z= 0\ ,
}\label{eqcw1}
\end{equation}

\noindent
with general solutions \cite{bocaletti01}, \cite{bakulin}

\begin{equation}
\eqalign{
\hspace{-2cm}x ( t) =  x_0+2\ \frac{\dot y_0}{\omega_0}-3 \,\left(\frac{\dot x_0}{\omega_0}-2 y_0 \right) t\, -2\frac{\dot y_0}{\omega_0} \cos \omega_0 t +2 \left(2 \frac{\dot x_0}{\omega_0}-3 y_0\right)\sin \omega_0 t \cr
\hspace{-2cm}y (t)  =  2\left(2 y_0- \frac{\dot x_0}{\omega_0}\right)+\left(2 \frac{\dot x_0}{\omega_0}-3 y_0\right)\cos \omega_0 t +\frac{\dot y_0}{\omega_0} \sin \omega_0 t \cr
\hspace{-2cm}z (t)  =  z_0 \cos  \omega_0 t  +\frac{\dot z_0}{\omega_0} \sin  \omega_0 t \ ,
}
\end{equation}

\noindent
where $x_0,y_0,z_0$ and $\dot x_0,\dot y_0,\dot z_0$ are the initial positions and velocities respectively.

\noindent
Since $x(t)$ contains a term proportional to $t$, after some time the assumption $\rho \ll R_0$ is no more valid and the above approximation breaks down.
\noindent
However, the divergent term can be cancelled by choosing $\dot x_0=2 \omega_0 y_0$. For the LISA case, the constraints of rigid and bounded relative motions lead to the solutions \cite{dhu05}

\begin{equation}
\eqalign{
x_k (t)  =  -\rho_0 \sin \left[\omega_0 t+ \sigma_k \right]  \cr
y_k (t)  =  -\frac{1}{2} \rho_0 \cos \left[\omega_0 t+ \sigma_k \right] \cr
z_k (t)  =  - \frac{\sqrt{3}}{2}\,\rho_0 \cos \left[\omega_0 t+ \sigma_k \right]\ ,
}\label{sol_hcw1}
\end{equation}

\noindent
where $\sigma_k =(1-k)\ 2 \pi /3$ and $\,k=1,2,3$ is a label which enumerates the LISA spacecrafts.

\subsection{Rescaling and expansion of the HCW equations}\label{normalization}
For our calculation we rewrite \eref{eqcw1} by means of the coordinate transformations $\hat x = x/ \rho_0, \hat y =y/ \rho_0, \hat z = z/ \rho_0$ and $\hat t = \omega_0 t$.

\begin{equation}
\eqalign{
{\hat x}'' - 2 {\hat y}' = 0 \cr
{\hat y}'' + 2 {\hat x}' -3 {\hat y} = 0\cr
{\hat z}'' + {\hat z}= 0 }
\label{eqcw1norm}
\end{equation}

\noindent
with the notation $\ '=d/d \hat t$. We will refer to \eref{eqcw1norm} as the HCW1 equations. We will show that the higher order terms in the expansions are negligible for calculating the perturbation of the LISA rigid and bounded orbits.

\noindent
It is worth noticing that the right-hand side of \eref{eqcw1norm} is zero only at the first order in the force expansion. In general, the right-hand side of \eref{eqcw1norm} is a polynomial of degree $n$, where $n$ is the order of the expansion \cite{nayak06}. In our case, the order of magnitude of the neglected terms is $\varepsilon_{HCW}=\rho_0/R_0 \simeq 1.9 \times 10^{-2}$.

\section{Gravitational near field of the EM system}\label{ems}

In our model, the EM system is constituted by 2 point masses: $m_1 \simeq 3.0 \times 10^{-6} M_\odot$ (Earth) and $m_2 \simeq 3.7 \times 10^{-8} M_\odot$ (Moon) located at a constant distance $l \simeq 2.57 \times 10^{-3}\ AU$ and describing a circular orbit with angular velocity $\omega_M=2\pi/P_M$, where $P_M \simeq 29.53\ d$ is the synodic month, around their common barycenter.
We assume that the barycenter of the EM system makes a circular orbit with radius $1 \ AU$ around the Sun, i.e. we neglect the eccentricity $\simeq 0.0167$ of the Earth orbit around the Sun.

\noindent
In addition, we disregard the eccentricity of the Moon orbit around the Earth ($\simeq 0.054$), its inclination to the ecliptic plane ($\simeq 5.14^\circ$), the motion of the perigee of the Moon ($\simeq 8.85\ yr$) and the precession of the Moon orbit plane ($\simeq 18.03\ yr$) \cite{abhyankar}, \cite{murray}.

\noindent
We consider a non rotating reference frame $\{X,Y,Z\}$ centered on the EM system barycenter, with the $X$ axis along the line joining Earth and Moon at $t=0$, and the $Z$ axis perpendicular to the ecliptic plane (see \Fref{cw}).
The gravitational potential due to the Earth and the Moon at a point $(X,Y,Z)$ is given by

\begin{eqnarray*}\label{eq_U}
\hspace*{-1.5cm}
U=-\frac{G
   m_1}{\sqrt{\left[X-\frac{
   m_2}{m_1+m_2} l \cos(\omega_M t) \right]^2+\left[Y-\frac{ m_2
   }{m_1+m_2} l \sin(\omega_M t)\right]^2+Z^2}}+\\
   \hspace*{-0.5cm}
   -\frac{G {m_2}}{\sqrt{\left[X+\frac{
   m_1}{m_1+m_2} l \cos(\omega_M t) \right]^2+\left[Y+\frac{ m_1
   }{m_1+m_2}l \sin(\omega_M t) \right]^2+Z^2}}\ .\\
\end{eqnarray*}

\subsection {Multipole expansion of the EM gravitational potential}

We are interested in the effect of the EM system on the LISA constellation, located at a distance $r_g$. The size $l$ of the EM system is small relatively to $r_g$, $l /r_g=7\times 10^{-3}$. We therefore expand the total potential in series of $\varepsilon_M=l /r_g$.

\begin{equation}\label{useries}
U(X,Y,Z,t) \equiv \sum_{n=0}^\infty \frac{1}{n!}\ \varepsilon_{EM}^n \ U_n (X,Y,Z,t)\ ,
\end{equation}

\noindent
where $U_n$ are the well known multipole terms.

\noindent
At the zeroth order we have the {\em monopole term}

\begin{equation}\label{terminemonopolo}
U_0 (X,Y,Z)=-\frac{G ({m_1}+{m_2})}{\sqrt{X^2+Y^2+Z^2}}\ .
\end{equation}
\noindent
The first order term, the {\em dipole} term, is equal to zero, due to the conservation of the linear momentum. The second order term, the {\em quadrupole term}, is

\begin{equation}\label{terminequadupolo}
\hspace*{-2.5cm}U_2(X,Y,Z,t)=\frac{m_1 m_2}{m_1+m_2}\ r_g^2\ \left[ \frac{X^2+Y^2+Z^2-3 \left[ X \cos \omega_M t +Y \sin \omega_M t  \right] ^2}{ (X^2+Y^2+Z^2)^{5/2}} \right]
\end{equation}
\noindent
and so on.

\noindent
Each $U_n$ term contains sinusoidal terms as $\sin k \omega_M t\ $ and $\cos k \omega_M t $, with $k=0,2,\dots, n$ for {\em even} $n$ and $k=1,3,\dots, n$ for {\em odd} $n$.

\noindent
Defining $\hat \omega_M=\omega_M / \omega_0 \simeq 12.3687$ and operating the substitution $X=x+x_g,\ Y=y+y_g,\ Z=z$ in all $U_n$, we obtain the EM potential in the HCW frame. The corresponding force per unit mass is

\[
\vect{F}_n=-\frac{1}{n!}\ \varepsilon_{EM}^n\, \nabla U_n (x,y,z,t)\ .
\]

\noindent
The resulting functions $\vect{F}_n(x,y,z,t)$ are still too complex to be treated analytically, but it is worth noticing that the values of the coordinates range within $-\rho_0$ and $+\rho_0$, while $r_g$ is about 20 times larger ($\rho_0/r_g = \varepsilon_L \simeq 5.5\times 10^{-2}$).
We can therefore expand $\vect{F}_n$ in terms of $x/r_g,y/r_g,z/r_g$ around the origin of the HCW frame

\begin{equation}\label{eqfin0}
\hspace{-2cm} \vect{F}_n( x, y, z, t)= \sum_{m=0}^\infty \frac{\varepsilon_L^m}{m!}\ \left[ \left(x \frac{\partial}{\partial \xi}+ y \frac{\partial}{\partial \eta}+ z \frac{\partial}{\partial \zeta} \right)^m \vect{F}_n(\xi,\eta,\zeta, t)\right]_{\xi=0,\eta=0,\zeta=0} \ .
\end{equation}

\noindent
The above formula can be expressed in a more useful way (after the rescaling)

\begin{equation}\label{eqfin}
\hspace{-2cm} \vect{F}_n(\hat x, \hat y, \hat z, \hat t)=\sum_{i=0}^\infty \frac{1}{i\ !} \sum_{j=0}^\infty \frac{1}{j\ !}\ \sum_{k=0}^\infty\ \frac{1}{k\ !}\ \varepsilon_L^{i+j+k}\ \vect{a}_{n,ijk}(\hat t)\ \hat{x}^i \hat{y}^j \hat{z}^k\ ,
\end{equation}

\noindent
where the $\vect{a}_{n,ijk}(\hat t)$ are

\begin{equation}\label{albetgam}
\vect{a}_{n,ijk}(\hat t)=\sum_m \left[ \mbf {\alpha}_{n,ijk,m}\sin  m \hat \omega_M \hat t  +\mbf {\beta}_{n,ijk,m} \cos  m \hat \omega_M \hat t \, \right]
\end{equation}

\noindent
with $\mbf \alpha_{n,ijk,m},\mbf \beta_{n,ijk,m}$ numerical coefficients and $m=0,2,\dots, n$ for {\em even} $n$ and $m=1,3,\dots, n$ for {\em odd} $n$.
\noindent
Note that for {\em even} $n$, \eref{albetgam} contains a constant term (i.e. $\mbf \beta_{n,ijk,0}$) plus sinusoids, while for {\em odd} $n$ it presents only sinusoids ($\mbf \beta_{n,ijk,1}$ is multiplied by $\cos\,\hat \omega_M \hat t $).

\noindent
Let us estimate the intensity of the acceleration due to the EM system : the most important contribution is given by the monopole term ($n=0$), and we assume its value at the origin of the coordinates as an indicator of its intensity. This term is

\[
\varepsilon_0=\frac{G (m_1+m_2) }{\omega_0 ^2\, r_g^2\, \rho_0} \simeq 1.31 \times 10^{-3} \simeq 7 \times 10^{-2} \varepsilon_{HCW}
\]
\noindent
therefore the EM system influence is about the 7\% of the contribution of the Sun (\Sref{normalization}).

\noindent
At the first order in the coordinates the monopole term (normalized to $\varepsilon_0$) is

\begin{eqnarray}
\eqalign{
\hspace{-1cm}f_x=\frac{x_g}{r_g} - \hat x\ \frac{\rho_0 (r_g^2 - 3 x_g^2)}{r_g^3} + \hat y\ \frac{3 \rho_0 x_g y_g}{r_g^3}\\
\hspace{-1cm}f_y= \frac{y_g}{r_g} +\hat x\ \frac{3 \rho_0 x_g y_g}{r_g^3} -\hat y\ \frac{\rho_0 (r_g^2 - 3 y_g^2)}{r_g^3}\\
\hspace{-1cm}f_z=- \hat z\ \frac{\rho_0}{r_g}\ , \\
}
\label{numericmonopole}
\end{eqnarray}
\noindent
where $\hat x, \hat y, \hat z$ are the scaled coordinates.

\noindent
The most important multipole term is the quadrupole ($n=2$) which is mostly constituted by the zeroth order term of its expansion in spatial coordinates, or equivalently, the force at the origin. This term is periodic with period $P_M/2=\pi/\omega_M$, and its mean value is

\[
\hspace{-1cm}
 \varepsilon_2=\frac{3}{4}\ \frac{G \,l^2}{\omega_0 ^2\ r_g^4\ \rho_0}\ \frac{m_1 m_2}{m_1+m_2} \simeq 6.48 \times 10^{-10} \ \simeq \ 5 \times 10^{-7}\ \varepsilon_0\ .
\]

\noindent
The monopole force has an identical time-independent polynomial structure as the expansion in coordinates of the Sun force. The difference is that this latter starts with second order terms \cite{nayak06}, while the EM monopole presents also linear terms and a constant term. Moreover, the monopole force is quite smaller than the Sun force.

\noindent
The multipole force with {\em even} $n$ also contains time-independent terms but the largest ones (corresponding to $n=2$) have order of magnitude $\simeq \varepsilon_2 \ll \varepsilon_{HCW}$.

\noindent
We can separate the total force in a more useful way:

\begin{enumerate}
\item A component $\vect{F_0}(\hat x,\hat y,\hat z)$ independent of time due to the terms of the Sun force ($\propto \varepsilon_{HCW}$) \cite{nayak06}, plus the EM system monopole ($\propto \varepsilon_0$, \eref{eqfin} with $n=0$) plus terms contained in the $n$-even multipole expansion ($\propto \varepsilon_2$, \eref{albetgam} with $m=0$);

\noindent
\item A component $\vect{F_1}(\hat t)$ that depends {\em only} on time due to the EM system multipole terms ($\propto \varepsilon_2$, \eref{eqfin} with $i,j,k=0$ and $n \geq2$). Physically this is the force at the origin of the HCW system;

\item A component $\vect{F_2}(\hat x,\hat y,\hat z,\hat t) \propto \varepsilon_2$ that depends on both time and coordinates (all other cases: \eref{eqfin} with $n \geq2$ and $m \neq 0$).
\end{enumerate}

\section{Perturbation of the LISA orbits due to EM system}\label{perturb}
If we indicate with $\vect{r}_{0i}(t)$ the unperturbed trajectory of the $i^{th}$ LISA test mass ($i=1,2,3$) and with $\vect{r}_{1i}(t)$ its small perturbation, the difference between two perturbed trajectories is simply
\[
\Delta \vect{r}_{ij}(t) \equiv \vect{r}_i(t)-\vect{r}_j(t) = \Delta \vect{r}_{0ij}(t) + \Delta \vect{r}_{1ij}(t)\ ,
\]

\noindent
where $\Delta \vect{r}_{0ij}(t)= \vect{r}_{0i}(t)-\vect{r}_{0j}(t)$, $\Delta \vect{r}_{1ij}(t)= \vect{r}_{1i}(t)-\vect{r}_{1j}(t) $.

\noindent
The perturbation of the relative displacement between the pair $i,j$ of LISA test masses can be written as
\begin{equation}\label{deltaLapprox}
\Delta L_{ij}(t) \simeq \frac{\Delta \vect{r}_{0ij} \cdot \Delta \vect{r}_{1ij}}{L_{0ij}}\ ,
\end{equation}

\noindent
where $L_{0ij}=\vert \Delta \vect{r}_{0ij} \vert$ is the distance between $i$ and $j$ test masses in the unperturbed case, and $\vert \vect{r}_{1i} \vert \ll \vert \vect{r}_{0i} \vert$.
We also define the perturbation to the differential distances between each pair of LISA as
\begin{equation}\label{deltadeltaLapprox}
\delta L_{ijk}(t) \equiv\Delta L_{ij}(t)-\Delta L_{jk}(t)
\end{equation}

\noindent
which represents the variation of $L_{ij}(t)-L_{jk}(t)$ due to a small perturbation and relates directly to the LISA sensitivity curve.

\subsection{Effects of the EM monopole}\label{emsmonopole}

At the first order, under the effect of the Sun, the LISA motion is described by \eref{sol_hcw1} and the EM monopole perturbation is contained in the time-independent force per unit mass $\vect{F_0}$.

\noindent
We write the motion of the $k$-spacecraft under the effect of
$\vect{F_0}$ in the following form: $\hat \vect{r}_{k}=\hat
\vect{r}_{0k}+\varepsilon_{HCW}\hat \vect{r}_{1k}+ \varepsilon_0 \hat
\vect{r}_{2k}$, where $\hat \vect{r}_{0k}$ is the unperturbed motion
(\eref{sol_hcw1}, rescaled), and $\hat \vect{r}_{1k}$ and $\hat
\vect{r}_{2k}$ are the perturbations due to the Sun force terms
\cite{nayak06} and the EM monopole, respectively. Being
$\varepsilon_{HCW}^2 \ll \varepsilon_0$, to calculate $\hat
\vect{r}_{2k}$ it is not necessary to know $\hat \vect{r}_{1k}$, \cite{Povoleri06}.

\noindent
The equations for $\hat \vect{r}_{2k}=(\hat x_{2k},\hat y_{2k},\hat z_{2k})$ are \cite{dhu08}, \cite{Povoleri06}
and \cite{dhu09}

\begin{equation}
\hspace{-1.5cm}
\eqalign{
{\hat x_{2k}}'' - 2 {\hat y_{2k}}' = f_x(\hat x_{0k},\hat y_{0k},\hat z_{0k}) \cr
{\hat y_{2k}}'' + 2 {\hat x_{2k}}' -3 {\hat y_{2k}} = f_y(\hat x_{0k},\hat y_{0k},\hat z_{0k})\cr
{\hat z_{2k}}'' + {\hat z_{2k}}= f_z(\hat x_{0k},\hat y_{0k},\hat z_{0k})\ , }
\label{eqmonopole}
\end{equation}

\noindent
where ($f_x,f_y,f_z$) are evaluated along the trajectory $\hat \vect{r}_{0k}$, using \eref{numericmonopole}.

\noindent
The solution can be written as

\begin{equation}
\eqalign{
\hspace{-1.5cm} \hat x_{2k}= c_{1,k} + c_{2,k} \hat t + c_{3,k} \hat t^2 + (c_{4,k}+c_{5,k} \hat t) \sin \hat t + (c_{6,k}+c_{7,k} \hat t) \cos \hat t \cr
\hspace{-1.5cm}\hat y_{2k} = c_{8,k} + c_{9,k} \hat t + (c_{10,k}+c_{11,k} \hat t) \sin \hat t + (c_{12,k}+c_{13,k} \hat t) \cos \hat t \cr
\hspace{-1.5cm}\hat z_{2k} = (c_{14,k}+c_{15,k} \hat t) \sin \hat t + (c_{16,k}+c_{17,k} \hat t) \cos \hat t\ ,
}
\label{solmonopole}
\end{equation}

\noindent
where $c_{1,k} \dots c_{17,k}$ are constants that depend on the initial conditions and on the geometric parameters ($\rho_0,x_g,y_g$).

\noindent
The solution contains terms $\propto \hat t$ and $\propto \hat t^2$, moreover there are also {\em mixed perturbations} as $\hat t \sin \hat t$, i.e. perturbations increases with time.

\begin{figure*}[h!]
\centering
\includegraphics[width=.49\columnwidth,height=.49\columnwidth]{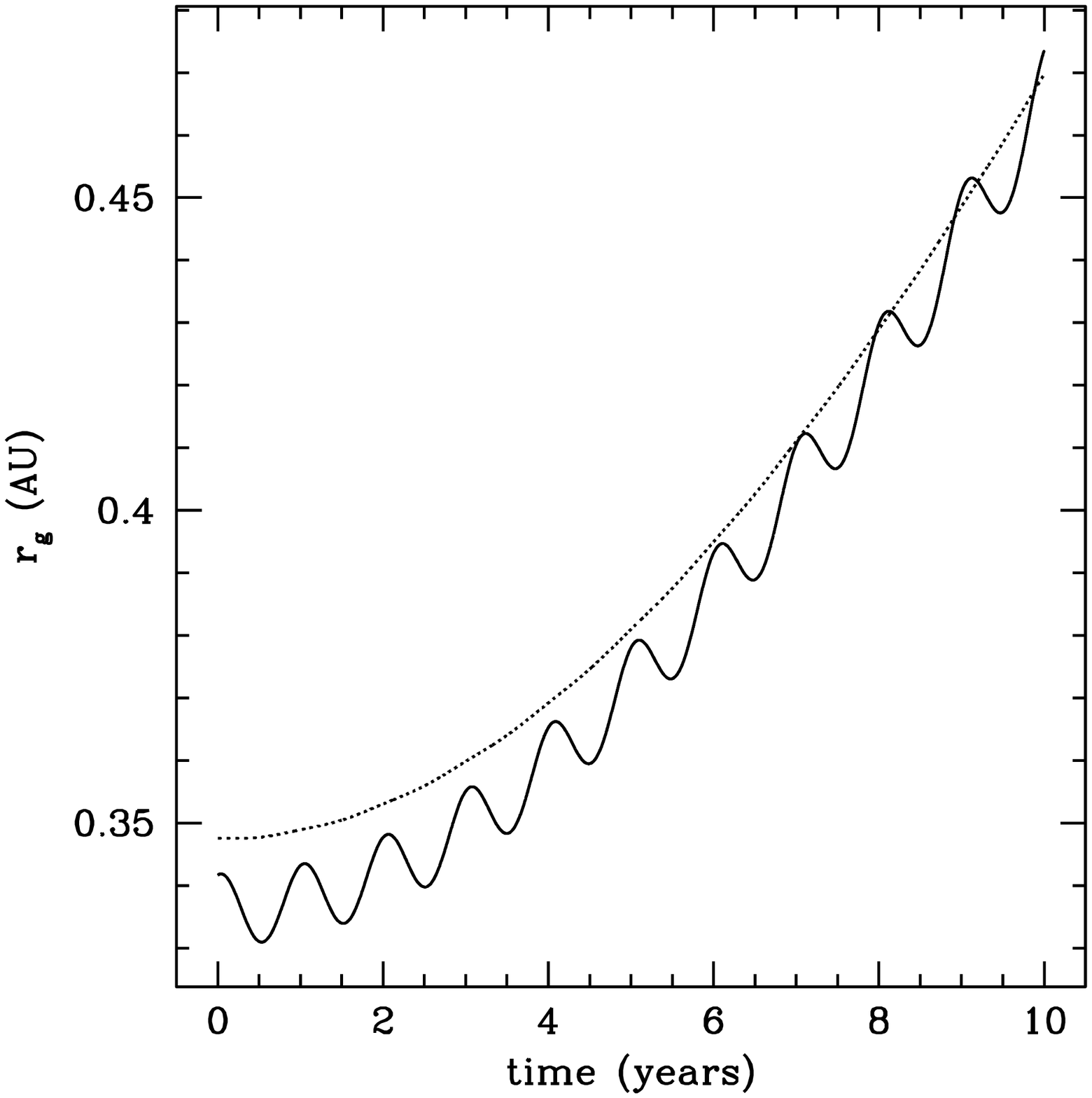}
\includegraphics[width=.49\columnwidth,height=.49\columnwidth]{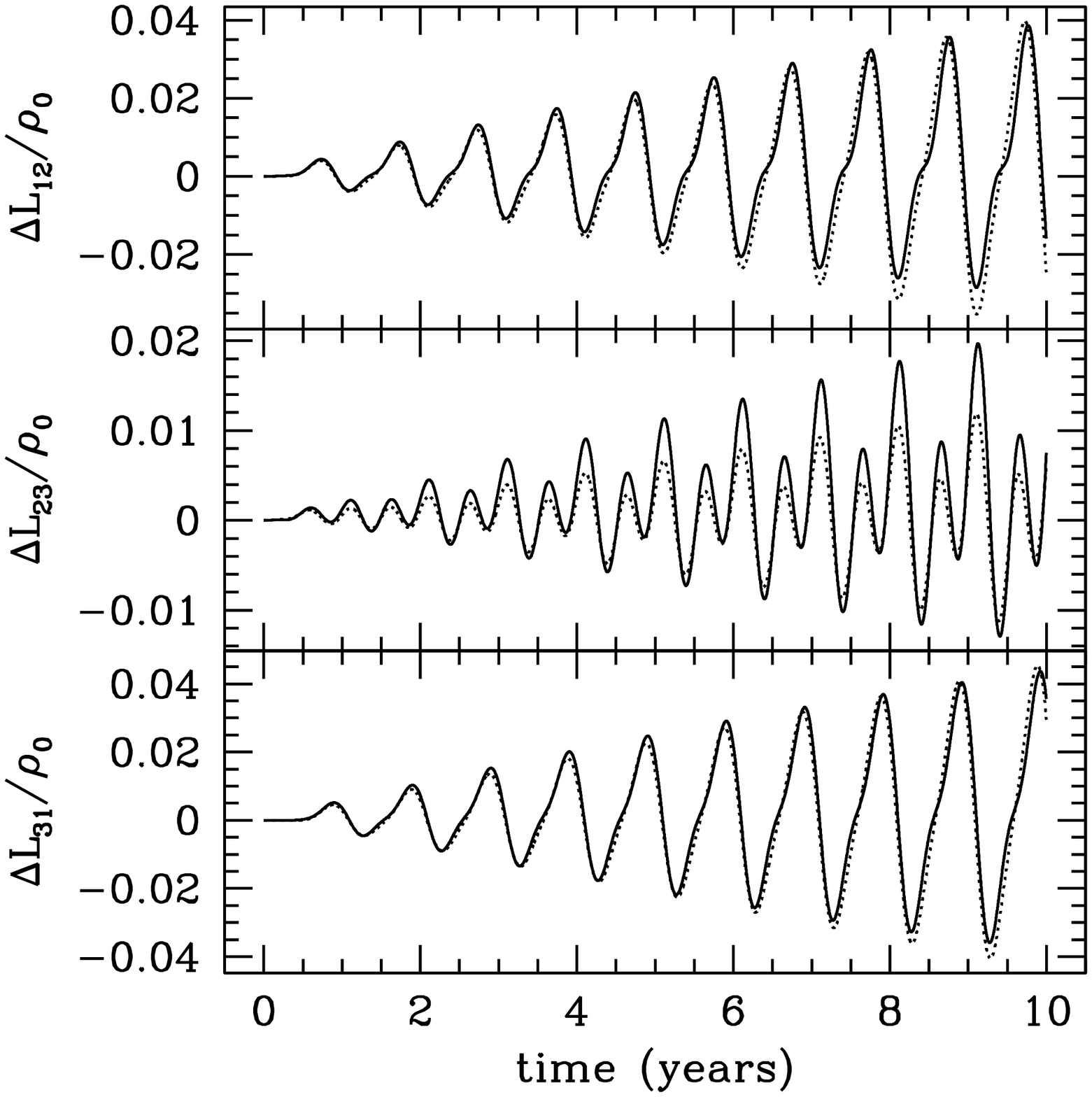}
\caption{{\emph Left panel}: Evolution of $r_g$ during 10 years: real (solid line) and simplified case (Earth describes a circular orbit around the Sun). The periodic component in the solid line is due to the eccentricity of the Earth orbit, while the trend is due to the EM system influence. {\emph Right panel}: perturbation of the LISA arms length due to the EM system monopole effect (rescaled): comparison between the numerical and the analytical calculation (solid and dashed lines, respectively).}
\label{rgandbreath}
\end{figure*}

\noindent
In particular, the coefficient of $\hat t^2$, $c_{3,k}=-3\, x_g /(2\, r_g)$ is positive number and the same for all spacecrafts. This means that the {\em entire} constellation is ''pushed away'' by the EM system ($r_g$ increases with time).
The variation of the LISA arms length can be calculated using \eref{sol_hcw1}, \eref{solmonopole} and \eref{deltaLapprox} and the result is represented in \Fref{rgandbreath} (right panel).

\noindent
The indefinite increasing of the perturbation \eref{solmonopole} is not physical because the perturbative regime would not be valid anymore, after few years. This is a direct consequence of the force linearization: the terms proportional to $\hat t \sin \hat t$ and $\hat t \cos \hat t$ are first-order terms of the real, {\em bounded}, solution for small $\hat t$. Using the F77 code \cite{tj}, we found that the perturbative pulls lead to a complete dismembering of the constellation and a successive recombination will occur after several tens of thousand year. In this scenario, the distance of each spacecraft from the HCW frame origin ranges from zero to $2\ AU$. Such motion is not a solution of the HCW equations, which are valid only if $\rho \ll R_0$. An all-time valid solution to our perturbative problem can be obtained applying the Lindstedt-Poincar\'e method \cite{gomez05}. However, as we are interested in the LISA motion during a few complete orbits, the Lindstedt-Poincar\'e method is not necessary.

\noindent
The increase of $r_g$ is shown in \Fref{rgandbreath} (left panel), where its time evolution is represented, during the hypothetical first 10 years of the mission.
The perturbation $\Delta L_{ij}$ to the relative motion between the pair $i,j$ of LISA test masses due to the monopole perturbation, calculated both analytically (dashed line) and via numerical integration \cite{tj} (solid line), is plotted on the right panel of \Fref{rgandbreath}. It is worth noticing the good agreement during the 10 years of the LISA mission.

\subsection{Effects of the EM multipoles}

We now search for a perturbative solution to the HCW1 equations in presence of $n \geq 2$ multipole terms. The intensity of this force is of the order of $\varepsilon_2$.
\noindent
We have already shown that the multipole force is composed by a periodic and a polynomial component independent of time.
The polynomial component is not important, as it can be added to the EM system monopole expansion and solved. The solution has the same structure as \eref{solmonopole}, with different coefficients (rescaled of a factor $\varepsilon_2/\varepsilon_0$), and the motion described in the previous section is therefore a very good approximation of the EM system polynomial component influence.

\subsubsection{$\vect{F_1}(t)$: periodic solutions in $n \hat \omega_M t$.}\label{f1calculus}

At the zeroth order ($i,j,k=0$) the multipole force does not depend on the coordinates. Therefore, for each order $n$ we have periodic terms in $n \hat \omega_M \hat t$. These terms represent the ''pure'' oscillations of a test mass due to the EM system that are not involved with $\omega_0$ harmonics.
\noindent
The solutions relative to these frequencies are equal for each test mass, being independent of its position (we can interpret this as a common motion). Since we are interested in the relative motion of the LISA satellites, we know {\em a priori} that these terms are subtracted when one measures the distance between two satellites.

\noindent
The equations to be solved are

\begin{equation}
\hspace{-1cm}
\eqalign{
 \hat x'' -2 \hat y'            = \sum_n \left[ a_{xn} \sin n \hat \omega_M \hat t +b_{xn} \cos n \hat \omega_M \hat t\, \right]\cr
 \hat y'' +2 \hat x' -3 \hat y  = \sum_n \left[ a_{yn} \sin n \hat \omega_M \hat t +b_{yn} \cos n \hat \omega_M \hat t\, \right]\cr
 \hat z'' + \hat z              = \sum_n \left[ a_{zn} \sin n \hat \omega_M \hat t +b_{zn} \cos n \hat \omega_M \hat t\, \right]\ ,
}
\label{hcw1xf0}
\end{equation}

\noindent
where $a_{xn,yn,zn}, b_{xn,yn,zn}$ are constants. Being $n \hat \omega_M \neq 1$ for each $n$, particular solutions can be written as

\begin{equation}
\hspace{-1cm}
\eqalign{
\hat x  = \sum_n  a'_{xn} \sin n \hat \omega_M \hat t +b'_{xn} \cos n \hat \omega_M \hat t \cr
\hat y  = \sum_n  a'_{yn} \sin n \hat \omega_M \hat t +b'_{yn} \cos n \hat \omega_M \hat t \cr
\hat z  = \sum_n  a'_{zn} \sin n \hat \omega_M \hat t +b'_{zn} \cos n \hat \omega_M \hat t
}
\end{equation}
\noindent
and the corresponding coefficients are

\begin{eqnarray}
\eqalign{
\hspace{-1cm}
a'_{xn}= -\frac{2 b_{yn} n \hat \omega_M+a_{xn} (3+n \hat \omega_M^2)}{n \hat \omega_M^2 (n \hat \omega_M^2-1)} \qquad &
 b'_{xn}= -\frac{-2 a_{yn} n \hat \omega_M+b_{xn} (3+n \hat \omega_M^2)}{n \hat \omega_M^2 (n \hat \omega_M^2-1)} \\
 \hspace{-1cm} a'_{yn}=\frac{2 b_{xn}-a_{yn} n \hat \omega_M}{n \hat \omega_M ( n \hat \omega_M^2-1)} \qquad &
b'_{yn}=\frac{-2 a_{xn}-b_{yn} n \hat \omega_M}{n \hat \omega_M ( n \hat \omega_M^2-1)} \\
\hspace{-1cm}  a'_{zn}=-\frac{{a_{zn}} }{n \hat \omega_M^2-1} \qquad & b'_{zn}=-\frac{{b_{zn}} }{n \hat \omega_M^2-1}\ .
}
\end{eqnarray}

\noindent
Inserting the numerical values, it comes out that the most important contribution is due to the $2\, \hat \omega_M$ frequency and it corresponds to an amplitude of about $1~cm$ for $x,y$ coordinates, while the coefficients $a'_{zn},b'_{zn}$ are all equal to zero.

\noindent
In \Tref{amplf1} we report the coefficients $a'_{xn,yn}, b'_{xn,yn}$ of the fluctuations in meters (i.e. multiplied by $\rho_0=2.89\times10^9\ m $), relatively to the first six harmonics of the fundamental frequency $\hat \omega_M$.

\begin{table}[!]
\caption{\label{amplf1} Rescaled coefficients
$\rho_0\, a'_{xn,yn}, \rho_0\, b'_{xn,yn}$
of the sinusoidal terms with frequency $n \hat \omega_M$ with $1 \leq n \leq 6$. }
\resizebox{0.7\columnwidth}{!}{
\begin{tabular}{@{}*{6}{l}}

\br
$n$ & Frequency & $\rho_0\, a'_{xn}$ & $\rho_0\, b'_{xn}$ & $\rho_0\, a'_{yn}$ & $\rho_0\, b'_{yn}$ \\
          &  [Hz]     & [m] & [m] & [m] & [m] \\
\mr					   							
 1 & 3.920$\times  10^{-7}$ & -4.5$\times 10^{-5}$   & -1.8$\times 10^{-4}$   & \ 6.5$\times 10^{-5}$  & -4.4$ \times 10^{-5}$   \\
 2 & 7.840$\times  10^{-7}$ & \ 4.4$\times 10^{-3}$  &\  8.5$\times 10^{-3}$  & -5.7$\times 10^{-3}$   & \ 3.9$ \times 10^{-3}$  \\
 3 & 1.176$\times  10^{-6}$ & -2.1$\times 10^{-5}$   & -2.7$\times 10^{-5}$   & \ 1.9$\times 10^{-5}$  & -1.8$ \times 10^{-5}$   \\
 4 & 1.568$\times  10^{-6}$ & \ 1.1$\times 10^{-7}$  &\ 1.0$\times 10^{-7}$   & -7.6$\times 10^{-8}$   &\ 9.8$  \times 10^{-8}$  \\
 5 & 1.960$\times  10^{-6}$ & -6.4$\times 10^{-10}$  & -3.9$\times 10^{-10}$  & \ 3.0$\times 10^{-10}$ & -5.7$ \times 10^{-10}$  \\
 6 & 2.352$\times  10^{-6}$ & \ 3.8$\times 10^{-12}$ & \ 1.5$\times 10^{-12}$ & -1.1$\times 10^{-12}$  & \ 3.4$ \times 10^{-12}$ \\
\br
\end{tabular}
}
\end{table}

\subsubsection{$\mathbf{F_2}(x,y,z,t)$: periodic solutions in $(n \hat \omega_M\pm m)\, \hat t$.}\label{f2calculus}
In \Sref{f1calculus} we showed that the $\vect{F}_1$ term corresponds to the particular solutions of \eref{hcw1xf0} and that $\vect{F}_1$ does not affect $\Delta L_{ij}$ being independent of coordinates.

\noindent
The motion associated with the coordinate--dependent term $\vect{F}_2$, will be different between each pair of test masses, and so the relative displacements $\Delta L_{ij}$ will be different from zero.

\noindent
The most direct approach to solve the equation of motion is to write HCW1 equations with $\vect{F}_2$ evaluated along the unperturbed trajectories given by \eref{sol_hcw1}.
Thus, we have obtained only the amplitudes relative to frequencies $(n \hat \omega_M \pm 1)$, which represent the main effect of $\vect{F}_2$. In order to have the complete spectrum $(n \hat \omega_M \pm m)$ we should consider the solutions of HCW with the complete expansion of the Sun force per unit mass (\Sref{normalization}).

\noindent
The solution can be written as sum of sinusoidal terms with frequencies $(n \hat \omega_M\pm 1)$. In addition, using \eref{deltaLapprox}, a similar relation can be written also for the perturbation $\Delta L_{ij}$ that in SI units reads

\begin{equation}\label{dlcoeff}
\hspace{-2cm}
\Delta L_{ij}(t) = \rho_0 \sum_n \sum_{m=-1}^1 \left[ a_{nm,ij}\sin   \omega_{nm} t  + b_{nm,ij} \cos   \omega_{nm} t \,  \right]\ ,
\end{equation}

\noindent
where $ \omega_{nm}= n \omega_M + 2 m \omega_0$, with $n=1,2, \dots ,+\infty$ and $m=-1,0,+1$.

\noindent
The quantity $\delta L_{ijk}$, which can be directly compared with the LISA sensitivity curve, reads

\begin{equation}\label{ddlcoeff}
\hspace*{-2cm}
\delta L_{ijk}(t) = \rho_0 \sum_n \sum_{m=-1}^1 \left[ a_{nm,ijk}\sin   \omega_{nm} t  + b_{nm,ijk} \cos   \omega_{nm} t\,   \right]\ ,
\end{equation}

\noindent
where $a_{nm,ijk}=a_{nm,ij}-a_{nm,jk}$ and $b_{nm,ijk}=b_{nm,ij}-b_{nm,jk}$.

\noindent
The coefficients of \eref{dlcoeff} and \eref{ddlcoeff} are reported in \Tref{spectrum1} and \Tref{spectrum4}, respectively, relatively to a certain number of frequencies.

\begin{figure*}[h!]
\centering
\includegraphics[width=.75\columnwidth,height=.7\columnwidth]{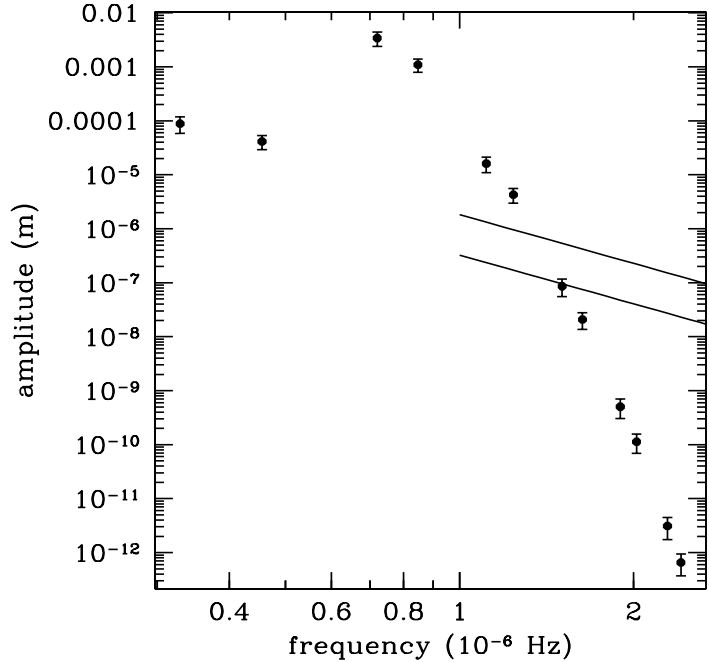}
\caption{
The filled circles give the amplitudes of the differential relative displacement $\delta L_{123}$ between the LISA test masses $1,2$ and $2,3$ induced by multipoles of the EM system; relative errors are also plotted. The two straight lines represent the LISA sensitivity curve for 12 days (upper straight line) and 1 year (lower straight line) of integration time.}
\label{rsn}
\end{figure*}

\noindent
In \Fref{rsn} we plot the $\delta L_{123}$ amplitudes (filled circles) superimposed to two LISA sensitivity curves
(straight lines) corresponding to integration times of 12 days (the upper one), period below which there should not be disturbances or if present should be removable and 1 year (the lower one), respectively.
The straight lines were obtained by extrapolating the LISA sensitivity curve down to $10^{-6}\ Hz$, as discussed in \cite{bender03}.

\noindent
Each amplitude is subject to time variations, because i) the Earth and the Moon orbits are not circular; ii) the orbital plane of the Moon is slightly inclined; and iii) the LISA constellation is not rigid.

\noindent
We estimated uncertainties for $x_g,y_g,l$ and $\rho_0$ by taking into account all these effects; in particular, we assumed the error relative to $\rho_0$ equal to the amplitude of the Sun induced breathing, i.e. $\simeq 2$\% of $\rho_0$ \cite{dhu05}.

\noindent
The resulting relative uncertainties of the $\delta L_{ijk}$ amplitudes are $\sim 30$\%.
\noindent

\begin{table}
\caption{\label{spectrum1} Coefficients relative to $\Delta L_{ij}(t)$. Amplitudes and frequencies are in SI units and the overall accuracy on amplitudes is $\simeq 30 \%$. }
\resizebox{0.99\columnwidth}{!}{
\begin{tabular}{@{}*{8}{l}}
\br
($n, m$) & Frequency & $a_{nm, 12}$ & $b_{nm, 12}$ & $a_{nm, 23}$ & $b_{nm, 23}$ & $a_{nm, 31}$ & $b_{nm, 31}$  \\
          &  [Hz]     & [m] & [m] & [m] & [m] & [m] & [m] \cr
\mr
(1,-1)   &   3.286$\times  10^{-7}$   &\ 5.0$\times 10^{-5}$ & -8.0$\times 10^{-6}$  &-1.8$\times 10^{-5}$ &\ 4.8$\times 10^{-5}$ & -3.2$\times 10^{-5}$ & -4.0$\times 10^{-5}$  \\
(1,\ 0)    &   3.920$\times  10^{-7}$   &\ 6.8$\times 10^{-6}$ & \ 1.6$\times 10^{-5}$   &\ 6.8$\times 10^{-6}$ &\ 1.6$\times 10^{-5}$ & \ 6.8$\times 10^{-6}$  &\ 1.6$\times 10^{-5}$   \\
(1,\ 1)    &   4.553$\times  10^{-7}$   &\ 1.9$\times 10^{-5}$ &  -1.4$\times 10^{-5}$  &\ 2.7$\times 10^{-6}$  &\ 2.4$\times 10^{-5}$  & -2.2$\times 10^{-5}$  & -9.5$\times 10^{-6}$         \\
(2,-1)   &   7.206$\times  10^{-7}$   &-2.1$\times 10^{-3}$ &  -2.1$\times 10^{-4}$  &\ 1.2$\times 10^{-3}$  & -1.7$\times 10^{-3}$  & \ 8.8$\times 10^{-4}$  &\ 1.9$\times 10^{-3}$                     \\
(2,\ 0)    &   7.839$\times  10^{-7}$   &-6.4$\times 10^{-4}$ &  -7.9$\times 10^{-4}$  &-6.4$\times 10^{-4}$  & -7.9$\times 10^{-4}$  & -6.4$\times 10^{-4}$  & -7.9$\times 10^{-4}$                 \\
(2,\ 1)    &   8.473$\times  10^{-7}$   &-3.9$\times 10^{-4}$ &  \ 5.0$\times 10^{-4}$  &-2.3$\times 10^{-4}$  & -5.9$\times 10^{-4}$  & \ 6.2$\times 10^{-4}$  &\ 9.1$\times 10^{-5}$                \\
(3,-1)   &   1.113$\times  10^{-6}$   &\ 8.9$\times 10^{-6}$ &  \ 2.6$\times 10^{-6}$  &-6.7$\times 10^{-6}$  &\ 6.4$\times 10^{-6}$  & -2.2$\times 10^{-6}$  & -9.0$\times 10^{-6}$      \\
(3,\ 0)    &   1.176$\times  10^{-6}$   &\ 3.6$\times 10^{-6}$ &  \ 3.0$\times 10^{-6}$  &\ 3.6$\times 10^{-6}$  &\ 3.0$\times 10^{-6}$  & \ 3.6$\times 10^{-6}$   &\ 3.0$\times 10^{-6}$     \\
(3,\ 1)    &   1.239$\times  10^{-6}$   &\ 1.1$\times 10^{-6}$ &  -2.2$\times 10^{-6}$  &\ 1.4$\times 10^{-6}$  &\ 2.0$\times 10^{-6}$  & -2.5$\times 10^{-6}$   &\ 1.6$\times 10^{-7}$      \\
(4,-1)   &   1.505$\times  10^{-6}$   &-4.4$\times 10^{-8}$ &  -2.2$\times 10^{-8}$  &\ 4.2$\times 10^{-8}$  & -2.7$\times 10^{-8}$  & \ 2.6$\times 10^{-9}$   &\ 5.0$\times 10^{-8}$ \\
(4,\ 0)    &   1.568$\times  10^{-6}$     &-2.3$\times 10^{-8}$ &  -1.3$\times 10^{-8}$  &-2.3$\times 10^{-8}$  & -1.3$\times 10^{-8}$  & -2.3$\times 10^{-8}$  &  -1.3$\times 10^{-8}$ \\
(4,\ 1)    &   1.631$\times  10^{-6}$   &-3.0$\times 10^{-9}$ &  \ 1.2$\times 10^{-8}$  &-8.5$\times 10^{-9}$  & -8.4$\times 10^{-9}$  & \ 1.2$\times 10^{-8}$  & -3.2$\times 10^{-9}$   \\
(5,-1)   &   1.896$\times  10^{-6}$   &\ 2.3$\times 10^{-10}$ & \ 1.8$\times 10^{-10}$ &-2.7$\times 10^{-10}$  &\ 1.1$\times 10^{-10}$  & \ 3.7$\times 10^{-11}$  & -2.9$\times 10^{-10}$ \\
(5,\ 0)    &   1.960$\times  10^{-6}$   & \ 1.5$\times 10^{-10}$ & \ 5.4$\times 10^{-11}$ &\ 1.5$\times 10^{-10}$  &\ 5.4$\times 10^{-11}$  &\ 1.5$\times 10^{-10}$  &\ 5.4$\times 10^{-11}$  \\
(5,\ 1)    &   2.023$\times  10^{-6}$   & \ 3.9$\times 10^{-12}$ & -6.5$\times 10^{-11}$&\ 5.4$\times 10^{-11}$  &\ 3.6$\times 10^{-11}$  & -5.8$\times 10^{-11}$  &\  2.9$\times 10^{-11}$ \\
(6,-1)   &   2.288$\times  10^{-6}$   &  -1.2$\times 10^{-12}$ & -1.3$\times 10^{-12}$&\ 1.7$\times 10^{-12}$  & -3.8$\times 10^{-13}$  & -5.4$\times 10^{-13}$  &\ 1.7$\times 10^{-12}$\\
(6,\ 0)    &   2.320$\times  10^{-6}$   &  -9.9$\times 10^{-13}$ & -1.7$\times 10^{-13}$ & -9.9$\times 10^{-13}$  & -1.7$\times 10^{-13}$  & -9.9$\times 10^{-13}$  & -1.7$\times 10^{-13}$ \\
(6,\ 1)    &   2.415$\times  10^{-6}$   & \ 5.0$\times 10^{-14}$ & \ 3.7$\times 10^{-13}$ & -3.5$\times 10^{-13}$  & -1.4$\times 10^{-13}$  &\ 3.0$\times 10^{-13}$  & -2.3$\times 10^{-13}$\\
\br
\end{tabular}
}
\end{table}

\begin{table}
\caption{\label{spectrum4} Coefficients relative to $\delta L_{ijk}(t)$.
 Amplitudes and frequencies are in SI units and the overall accuracy on amplitudes is $\simeq 30 \%$. }
\resizebox{0.99\columnwidth}{!}{
\begin{tabular}{@{}*{8}{l}}
\br
 ($n,m$)  & Frequency & $a_{nm,123}$ & $b_{nm,123}$ & $a_{nm,231}$ & $b_{nm,231}$ & $a_{nm,312}$ & $b_{nm,312}$  \\
          &  [Hz]     & [m ] & [m ] & [m ] & [m ] & [m ] & [m ] \\
\hline					   							
(1,-1)     &   3.286$\times 10^{-7}$ &\ 6.9 $\times 10^{-5}$ & -5.6 $\times 10^{-5}$ &\ 1.4 $\times 10^{-5}$ &\ 8.7 $\times 10^{-5}$ & -8.3 $\times 10^{-5}$ & -3.2 $\times 10^{-5}$\\
(1,\ 0)      &    3.920$\times 10^{-7}$ &  \;0 & \;0 & \;0 & \;0 & \;0 & \;0 \\
(1,\ 1)      &   4.553$\times 10^{-7}$ &\ 1.6 $\times 10^{-5}$ & -3.8$\times 10^{-5}$ &\ 2.5 $\times 10^{-5}$ &\ 3.3$\times 10^{-5}$ & -4.1 $\times 10^{-5}$ &\ 4.7$\times 10^{-6}$\\
(2,-1)     &   7.206$\times 10^{-7}$ &  -3.4 $\times 10^{-3}$ &\ 1.5 $\times 10^{-3}$ &\ 3.6 $\times 10^{-4}$ & -3.7 $\times 10^{-3}$ & \ 3.0$\times 10^{-3}$ &\ 2.1 $\times 10^{-3}$\\
(2,\ 0)      &   7.839$\times 10^{-7}$ &  \;0 & \;0 & \;0 & \;0 & \;0 & \;0 \\
(2,\ 1)      &   8.473$\times 10^{-7}$ &  -1.6$\times 10^{-4}$ &\ 1.1 $\times 10^{-3}$ & -8.6 $\times 10^{-4}$ & -6.8 $\times 10^{-4}$ & \ 1.0$\times 10^{-4}$ & -4.0$\times 10^{-4}$\\
(3,-1)     &   1.113$\times 10^{-6}$ &\ 1.6 $\times 10^{-5}$ & -3.8$\times 10^{-6}$ & -4.6$\times 10^{-6}$ &\ 1.5 $\times 10^{-5}$ & -1.1 $\times 10^{-5}$ & -1.2 $\times 10^{-5}$ \\
(3,\ 0)      &   1.176$\times 10^{-6}$ &  \;0 & \;0 & \;0 & \;0 & \;0 & \;0 \\
(3,\ 1)      &   1.239$\times 10^{-6}$ &  -2.8$\times 10^{-7}$ & -4.3$\times 10^{-6}$ &\ 3.8$\times 10^{-6}$ &\ 1.9$\times 10^{-6}$  & -3.6$\times 10^{-6}$ &\ 2.4$\times 10^{-6}$ \\
(4,-1)     &   1.505$\times 10^{-6}$ & -8.6$\times 10^{-8}$  &\ 4.5$\times 10^{-9}$ &\ 3.9$\times 10^{-8}$ & -7.7$\times 10^{-8}$ & \ 4.7$\times 10^{-8}$ &\ 7.2$\times 10^{-8}$ \\
(4,\ 0)      &   1.568$\times 10^{-6}$ &  \;0 & \;0 & \;0 & \;0 & \;0 & \;0 \\
(4,\ 1)      &   1.631$\times 10^{-6}$ &\ 5.5$\times 10^{-9}$  &\ 2.0$\times 10^{-8}$ & -2.0$\times 10^{-8}$ & -5.3$\times 10^{-9}$ & \ 1.5$\times 10^{-8}$ & -1.5$\times 10^{-8}$ \\
(5,-1)     &   1.896$\times 10^{-6}$ &\ 5.0$\times 10^{-10}$  &\ 6.5$\times 10^{-11}$ & -3.0$\times 10^{-10}$ &\ 4.0$\times 10^{-10}$ & -1.9$\times 10^{-10}$ & -4.6$\times 10^{-10}$ \\
(5,\ 0)      &   1.960$\times 10^{-6}$ &  \;0 & \;0 & \;0 & \;0 & \;0 & \;0 \\
(5,\ 1)      &   2.023$\times 10^{-6}$ & -5.0$\times 10^{-11}$ &  -1.0$\times 10^{-10}$ &\ 1.1$\times 10^{-10}$ &\ 6.8$\times 10^{-12}$ & -6.2$\times 10^{-11}$ &\ 9.4$\times 10^{-11}$ \\
(6,-1)     &   2.288$\times 10^{-6}$ & -3.0$\times 10^{-12}$  &  -9.4$\times 10^{-13}$ &\ 2.3$\times 10^{-12}$ & -2.1$\times 10^{-12}$ & \ 6.6$\times 10^{-13}$ &\ 3.0$\times 10^{-12}$ \\
(6,\ 0)      &   2.352$\times 10^{-6}$ &  \;0 & \;0 & \;0 & \;0 & \;0 & \;0 \\
(6,\ 1)      &   2.415$\times 10^{-6}$ &\ 4.0$\times 10^{-13}$  &\ 5.2$\times 10^{-13}$ & -6.5$\times 10^{-13}$ &\ 8.6$\times 10^{-14}$ & \ 2.5$\times 10^{-13}$ & -6.0$\times 10^{-13}$ \\
\br
\end{tabular}
}
\end{table}

\section{Effect of Venus and Jupiter on LISA motion}\label{planets}

\begin{figure*}[h!]
\centering
\includegraphics[width=.75\columnwidth,height=.65\columnwidth]{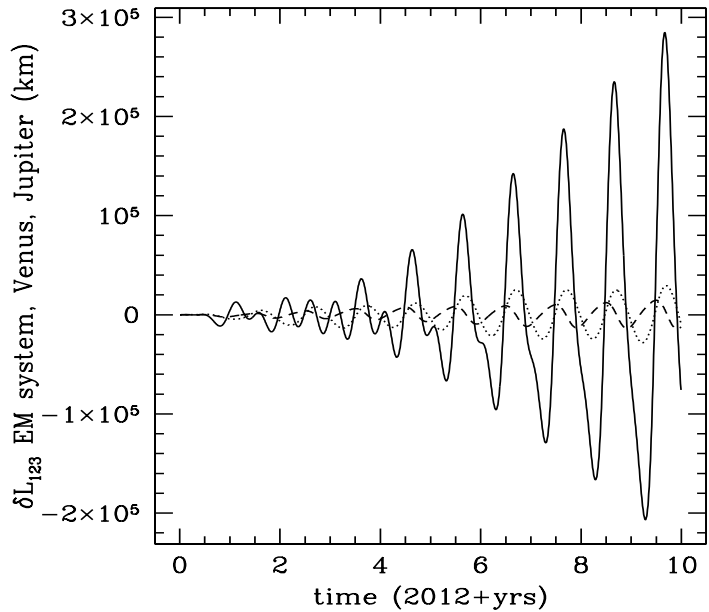}
\caption{\footnotesize
$\delta L_{123}$ perturbation (in km) due to EM system (solid line), Venus (dotted line) and Jupiter (dashed line). $\delta L_{123}$ perturbation (in km) due to monopole contribution of the EM system (solid line), Venus (dotted line) and Jupiter (dashed line). See \Sref{planets}.
}
\label{jupven}
\end{figure*}

\noindent
To make a comparison with the effect on the LISA motion due to EM system effect, we have numerically evaluated the effect of Venus, Jupiter and the EM system (its monopole contribution) using a F77 code implementing the algorithm described in \cite{tj}.
\noindent
In practice, we independently computed the LISA motion under the effect of Sun+Venus, Sun+Jupiter and Sun+EM system and we subtracted, to each of them, the unperturbed motion due to the Sun. In \Fref{jupven} we plot the $\delta L_{123}$ perturbations (in km) once the modulation due to the Sun has been subtracted. We have found that the monopole contribution of the EM system (solid line) is much larger than that of Venus (dotted line) and Jupiter (dashed line). \Fref{jupven} also shows the onset of the resonant effect of the EM system monopole after 2-3 years.

\section{Conclusions}\label{conclusion}
We calculated the relative motion of LISA test masses due to the effect of the EM system monopole and we found $\Delta L_{ij} \simeq 3\times 10^5 \ km$ after a period of 10 years.
We also found that the Jupiter and Venus influences are at least 10 times smaller than the EM system one.

\noindent
The perturbations of the differential relative motion of LISA test masses $\delta L_{ijk}$ are in the $10^{-6}$ to
$10^{-7}\ Hz$ decade.
This is a very low frequency range in which LISA residual acceleration noise may be much larger than extrapolated on grounds of known effects in \cite{bender03}.
However, it might not be completely hopeless to get an interesting sensitivity also at such low frequencies, provided that one will have to face the problem of i) loss of signal coherence over a time scale of one month; and ii) low frequency range calibration.

\noindent
As discussed in Pollack \cite{Pollack2004} it is possible to extract a signal from the LISA data, even in presence of disturbances. These latter arise due to environmental effects, such as cosmic rays induced by solar flares, and the telecommunication antenna which periodically has to be rotated. Pollack showed how these disturbances can be identified and subsequently removed from the data even at low frequencies. Assuming, for instance, that a disturbance appears every 19 days (see Table 6 in \cite{Pollack2004}) the resulting error in the signal frequency of $3 \times 10^{-6}\, Hz$ is only $\sim 1.3 \times 10^{-9}\, Hz$.
\noindent
Thus it should not be a problem to extend to such low frequencies the calibration from the verification binaries, and, by using their signals, ensure the continuity of data over time spans of many weeks.
\noindent
Still the signal from the EM system, as understood at the level of accuracy given in the present paper, can give a relevant additional crosscheck to such an extension of the calibration. It thus may help in improving our knowledge of the LISA acceleration noise at very low frequencies and contribute to extend to such low frequencies the capabilities of LISA.

\ack
We are indebted to Peter Bender for a critical reading of the initial
version of the manuscript, together with helpful
suggestions. We thank Oliver Jennrich and Gerard G\'omez, for useful discussions.
Mauro Sereno was supported in the early stages of this work by the Swiss National Science Foundation.

\section*{References}

\end{document}